\pdfoutput=1
\documentclass[11pt,aps,prd,a4paper,showpacs,showkeys,superscriptaddress, preprintnumbers,floatfix,nofootinbib,notitlepage]{revtex4-1}

\usepackage[toc,page]{appendix}
\usepackage{latexsym}
\usepackage{revsymb}
\usepackage{multirow}
\usepackage{color}
\usepackage[usenames,dvipsnames,svgnames,table]{xcolor}
\usepackage{booktabs}

\usepackage{graphicx}
\usepackage{epsfig}  
\usepackage{epsf}    
\usepackage{dcolumn}
\usepackage{bm}
\usepackage{textcomp}
\usepackage{float}
\usepackage[]{hyperref}
\usepackage{adjustbox}
\usepackage[utf8]{inputenc}
\usepackage[T1]{fontenc} 
\usepackage{url}
\usepackage{amsmath}
\usepackage{amssymb}

\usepackage{mathtools}
\usepackage{subfigure}
\usepackage{alphalph}
\usepackage{morefloats}
\usepackage{comment}
\usepackage{ulem}
\usepackage{booktabs}

\hypersetup{colorlinks=true,linkcolor=blue,citecolor=blue,urlcolor=blue}

\definecolor{gray}{rgb}{0.5,0.5,0.5}

\newcommand{\eps}{\varepsilon}

\newcommand{\AddrIPN}{%
Departamento de F\'{\i}sica, Centro de Investigaci\'on
  y de Estudios Avanzados del IPN,\\ Apartado Postal 14-740 07000,
  Ciudad de Mexico, Mexico}
  
\newcommand{\AddrIPhT}{Laboratorio Interdisciplinario de Investigación Tecnológica-LIIT, Tecnol\'ogico Nacional de M\'exico - ITS de Jerez, C.P. 99863 Zacatecas, Mexico.}

\newcommand{\AddrFC}{ {\it Departamento de Física, Facultad de Ciencias, Universidad Nacional Autónoma de México,
Apartado Postal 70-542, Ciudad de México 04510, Mexico
.}}

\begin{document}

\title{Generalized Neutrino Interactions: constraints and parametrizations}

\author{L. J. Flores} \email{ljflores@jerez.tecnm.mx} \affiliation{\AddrIPhT}

\author{O. G. Miranda} \email{omar.miranda@cinvestav.mx} \affiliation{\AddrIPN}

\author{G. Sanchez Garcia} \email{g.sanchez@ciencias.unam.mx} \affiliation{\AddrFC}


\begin{abstract}
Generalized neutrino interactions (GNI) are emerging as a convenient framework for describing effective scalar, vector, and tensor interactions. Such interactions arise naturally from extensions of the Standard Model that aim to explain neutrino properties and their mass origin. 
In this paper, we carefully study the two more common parametrizations for GNI and how to relate them.
This allows us to compare bounds obtained from CEvNS and deep-inelastic scattering under the same footing.
In addition, we present the current bounds from CEvNS measurements by COHERENT and compare them to those obtained from deep inelastic scattering on the same level. Our results focus on neutrino-quark interactions, and illustrate the complementarity between experiments working at different scales for GNI, showing that scalar interactions are better constrained by low-energy experiments like COHERENT, while tensor interactions are robustly constrained from deep inelastic scattering.
\end{abstract}

\maketitle

\section{Introduction}
\label{sec:introduction}

The experimental discovery of neutrino oscillations has represented an unambiguous evidence that neutrinos are massive particles~\cite{Kajita:2016cak, RevModPhys.88.030502}.  Hence,  neutrino states of a definite flavor observed through their interactions with matter fields are known to be a mixing of massive states. In the standard three-neutrino oscillation picture, these two different  basis are related through a $3\times 3$ unitary matrix which is parametrized by three mixing angles, $\theta_{12}, \theta_{13}, $ and $\theta_{23}$,  and one CP phase, $\delta$.  Currently, neutrino experiments are in a precision era in determining the parameters characterizing neutrino mixing~\cite{DayaBay:2022orm,T2K:2025wet, JUNO:2025gmd}, being the precise detemination of the CP-violating phase the next challenge to be faced by next-generation experiments such as DUNE~\cite{DUNE:2020lwj} and Hyper-K~\cite{Hyper-Kamiokande:2018ofw}.  A complete phenomenological review of the current status of neutrino oscillation parameters can be found in Ref. \cite{deSalas:2020pgw}. From the theoretical side, a plethora of models have been proposed in an attemp to explain the mechanism responsible for neutrino masses~\cite{Schechter:1980gr,Schechter:1981cv,Mohapatra:1986bd,Ma:2006km}. Depending on the particular model,  the introduction of  particle fields of different nature, and not present within the Standard Model (SM),  have been proposed either in the form of right handed neutrinos, scalar doublets and triplets, or new fermions~\cite{foot:1988aq, hirsch:2004he, Malinsky:2005bi, Grimus:2006nb}. Furthermore, some scenarios include Dark Matter (DM) candidates that can be tested either directly through nuclear recoils, or indirectly through their final state signals in particle colliders~\cite{ Candela:2024ljb,Lozano:2025tst}.

Examining all feasible existing neutrino mass models is a challenging task. However, we can optimize this work by using a parametrization encoding the main characteristics of wide families of models. For instance, several neutrino mass models induce slight corrections to the vector and axial couplings predicted within the SM framework.
These corrections can be included in a generic way through the so-called Non Standard Interactions (NSI), which can affect neutrino production, propagation, and detection. Under this parametrization, the new physics effects are contained in NSI parameters that have been widely tested through the scattering of neutrinos from electrons, nucleons, and nuclei~\cite{Barranco:2005yy,Ohlsson:2012kf,Miranda:2015dra,Farzan:2017xzy,Coloma:2023ixt}.
In its original formulation, the formalism of NSI is limited to the study of vector and axial current couplings.
However, there are also several scenarios of interest that predict additional scalar and tensor neutrino interactions. That is the case, for instance, of hypothetical mediators as leptoquarks, that can induce scalar corrections to neutrino interactions~\cite{Buchmuller:1986zs,Crivellin:2019dwb,DeRomeri:2023cjt}; on other hand, neutrino magnetic moments, tested in different neutrino detection experiments, correspond to tensor operators~\cite{AristizabalSierra:2020zod,DeRomeri:2024hvc,Demirci:2025poc}.
Interactions like these can be phenomenologically studied through the
formalism of Generalized Neutrino Interactions (GNI), which
parametrize effective scalar, pseudoscalar, vector, axial, and tensor
interactions, and have been tested with a variety of neutrino
experiments~\cite{Lindner:2016wff,
AristizabalSierra:2018eqm,Khan:2019jvr,
Han:2020pff,Escrihuela:2021mud, Chen:2021uuw,
Flores:2021kzl,DeRomeri:2022twg, Coloma:2022umy, DeRomeri:2022twg,
Escrihuela:2023sfb,
KATRIN:2024odq,Denton:2024upc,AristizabalSierra:2024nwf,Liao:2025hcs,Alpizar-Venegas:2025wor,Chattaraj:2025fvx,DeRomeri:2025csu,TEXONO:2025sub,Celestino-Ramirez:2025snn}.

Currently there are two widely known parametrizations for the GNI effects. In this paper, we provide a comprehensive guide to switch between these parametrizations such that the results of the analysis from different experiments can be compared under the same footage. The reminder of this paper is given as follows: in section \ref{sec:theory} we provide the theoretical framework of GNI under the two different paramterizations used in the literature. Then, in section \ref{sec:experiments} we describe the different neutrino interactions, and their corresponding experiments, that are used to constrain GNI parameters. In section \ref{sec:results} we show a review of the different constraints, and we give our conclusions in section \ref{sec:conclusions}.

\section{Generalized Neutrino Interactions}
\label{sec:theory}
There are two well-accepted parametrizations in the literature when considering the presence of GNI.  Here, we provide a complete guide to the mapping between them. We begin with what we will refer to as the `\textit{epsilon}' parametrization. In this case, the interaction Lagrangian that is added to the SM one reads
\begin{eqnarray}
\label{eq:epsilon}
\mathcal{L}^{eps}_{\mathrm{GNI}} = -\frac{G_F}{\sqrt{2}}
&& \left\{ \varepsilon_{\alpha\beta}^{qS}
\bigl(\overline{\nu}_\alpha (1-\gamma^5) \nu_\beta \bigr)
\bigl(\overline{q} q\bigr)
- \varepsilon_{\alpha\beta}^{qP}
\bigl(\overline{\nu}_\alpha (1-\gamma^5) \nu_\beta \bigr)
\bigl(\overline{q} \gamma^5 q\bigr) \right.  \nonumber \\
&& \left. +~\tilde{\varepsilon}_{\alpha\beta}^{qS}
\bigl(\overline{\nu}_\alpha (1+\gamma^5) \nu_\beta \bigr)
\bigl(\overline{q} q\bigr)
- \tilde{\varepsilon}_{\alpha\beta}^{qP}
\bigl(\overline{\nu}_\alpha (1+\gamma^5) \nu_\beta \bigr)
\bigl(\overline{q} \gamma^5 q\bigr) \right.  \nonumber \\
&& \left. +~\varepsilon_{\alpha\beta}^{qV}
\bigl(\overline{\nu}_\alpha \gamma^{\mu}(1-\gamma^5) \nu_\beta \bigr)
\bigl(\overline{q}\gamma^\mu q\bigr)
- \varepsilon_{\alpha\beta}^{qA}
\bigl(\overline{\nu}_\alpha \gamma^\mu(1-\gamma^5) \nu_\beta \bigr)
\bigl(\overline{q} \gamma^\mu\gamma^5 q\bigr) \right.  \nonumber \\
&& \left. +~\tilde{\varepsilon}_{\alpha\beta}^{qV}
\bigl(\overline{\nu}_\alpha \gamma^{\mu}(1+\gamma^5) \nu_\beta \bigr)
\bigl(\overline{q}\gamma^\mu q\bigr)
- \tilde{\varepsilon}_{\alpha\beta}^{qA}
\bigl(\overline{\nu}_\alpha \gamma^\mu(1+\gamma^5) \nu_\beta \bigr)
\bigl(\overline{q} \gamma^\mu\gamma^5 q\bigr) \right.  \nonumber \\
&&  \left. +~\varepsilon_{\alpha\beta}^{qT}
\bigl(\overline{\nu}_\alpha \sigma^{\mu \nu} (1-\gamma^5) \nu_\beta \bigr)
\bigl(\overline{q} \sigma_{\mu \nu}  (1-\gamma^5) q\bigr) \right.  \nonumber \\ 
&&  \left. +~\tilde{\varepsilon}_{\alpha\beta}^{qT}
\bigl(\overline{\nu}_\alpha \sigma^{\mu \nu} (1+\gamma^5) \nu_\beta \bigr)
\bigl(\overline{q} \sigma_{\mu \nu}  (1+\gamma^5) q\bigr) \right\}  ,
\label{eq:GNI}
\end{eqnarray}
where the indices $\alpha$ and $\beta$ run over the three neutrino flavors and  $\sigma^{\mu \nu} = \frac{i}{2} \left[ \gamma^\mu, \gamma^\nu \right]$\footnote{Some references write the tensor term proportional to $\bigl(\overline{\nu}_\alpha \sigma^{\mu \nu} P_L \nu_\beta \bigr)\bigl(\overline{q} \sigma_{\mu \nu}  q\bigr)$, which is equivalent to our parametrization since $\bigl(\overline{\nu}_\alpha \sigma^{\mu \nu} P_L \nu_\beta \bigr)\bigl(\overline{q} \sigma_{\mu \nu} P_R q\bigr) = 0$~\cite{Nieves:2003in}.}.  The above Lagrangian corresponds to interactions involving neutrinos and quarks of any generation. However, since we are interested in nucleon and  nucleus scattering, we will limit our analysis to the case where $q = u, d$. 
The first two lines in Eq. \eqref{eq:GNI} correspond to scalar interactions that, in general, may result from integrating out the degrees of freedom of heavy Higgs-like fields.  The third and fourth lines correspond to vector interactions that are similar in structure to the SM ones, with the difference that we allow for a non-universality of the interaction as well as transition between different flavor states. These interactions are commonly named as nonstandard interactions and have been widely studied in the literature~\cite{Ohlsson:2012kf,Miranda:2015dra,Farzan:2017xzy}. Finally, the last two lines in the above Lagrangian account for tensor interactions.
Notice that in general the parameters $ \tilde{\varepsilon}_{\alpha\beta}^{qX}\ $  in Eq. (\ref{eq:epsilon}) are nonzero.  However,  we should keep in mind that the terms proportional to these parameters involve interactions only  between right-handed states. 

A second parametrization accepted in the literature corresponds to an interaction Lagrangian of the form 
\begin{eqnarray}
\mathcal{L}^{CD}_{\mathrm{GNI}} = -\frac{G_F}{\sqrt{2}}
&& \left\{  \bigl(\overline{\nu}_\alpha \nu_\beta \bigr) 
\bigl(\overline{q} \bigl(  C_{\alpha\beta}^{qS} + i D_{\alpha\beta}^{qS}\gamma^{5}\bigr)q\bigr)  - \bigl(\overline{\nu}_\alpha \gamma^5  \nu_\beta \bigr) 
\bigl(\overline{q}\gamma^5 \bigl(  C_{\alpha\beta}^{qP} + i D_{\alpha\beta}^{qP}\gamma^{5}\bigr)q \bigr)   \right.  \nonumber \\
&& \left. +~\bigl(\overline{\nu}_\alpha \gamma^\mu \nu_\beta \bigr) \bigl(
\overline{q}\gamma_\mu \bigl( \mathbb{C}_{\alpha\beta}^{qV} - \mathbb{D}_{\alpha\beta}^{qV}\gamma^{5}\bigr)q \bigr) + \bigl(\overline{\nu}_\alpha \gamma^5\gamma^\mu  \nu_\beta \bigr)\bigl( 
\overline{q}\gamma^5\gamma_\mu \bigl(  \mathbb{C}_{\alpha\beta}^{qA} -  \mathbb{D}_{\alpha\beta}^{qA}\gamma^{5}\bigr)q \bigr) \right.  \nonumber \\
&&  \left. +~\bigl(\overline{\nu}_\alpha \sigma^{\mu\nu} \nu_\beta \bigr) 
\bigl(\overline{q}\sigma_{\mu\nu} \bigl(  C_{\alpha\beta}^{qT} + i D_{\alpha\beta}^{qT}\gamma^{5}\bigr)q\bigr)  \right\}  ,
\label{eq:GNI:D}
\end{eqnarray}
which we refer to as the `\textit{C}' parametrization. The curvy notation in the vector couplings is used to distinguish that, as written above, the Lagrangian in Eq. \eqref{eq:GNI:D} includes SM interactions.  To separate SM from new vector interactions, we make the change\footnote{Some references may differ by factors of $\sqrt{2}$. Historically, weak interactions questioned if only vector currents existed in nature~\cite{Lee:1956qn,Wu:1957my}. Although parity conservation was the main concern~\cite{Lee:1956qn}, all possible couplings were discussed. The same can be say about the $V-A$ theory~\cite{Sudarshan:1958vf}. Despite focusing in vector and axial couplings, all the couplings were considered in the original effective theory. It was only after the Standard Model that these effective couplings were predicted from the gauge principle, $C_V= 2g_v= 2 (t^3_L + 2Q \sin^2\theta_W)$ and $C_A= 2 g_A =2 t^3_L$ while all other couplings are taken to be zero. 
}
\begin{eqnarray}
&&\mathbb{C}^{qV}  = 2g_V^\nu g_V^q + C^{qV}  \nonumber \\
 && \mathbb{C}^{qA}  = 2g_A^\nu g_A^q +  C^{qA}   \nonumber \\
&& \mathbb{D}^{qV}  = -2g_V^\nu g_A^q + D^{qV}   \nonumber \\
&&  \mathbb{D}^{qA}  = -2g_A^\nu g_V^q +  D^{qA},  
\label{eq:GNI:conv}
\end{eqnarray}
where $g_V^\nu = g_A^\nu = 1/2$,  $g_A^u = -g_A^d = 1/2$,  $g_V^u = 1/2 - 4/3 \sin^2{\theta_W}$, and $g_V^d = -1/2 + 2/3 \sin^2{\theta_W}$ are the coupling constants defined within the SM. 

To convert between the two parametrizations in Eqs. \eqref{eq:GNI} and \eqref{eq:GNI:D}, we use the following relations
\begin{itemize}
\item[a)] Scalar parameters:
\begin{equation}
\varepsilon_{\alpha\beta}^{qS} = \frac{1}{2}\left ( C_{\alpha\beta}^{qS} + iD_{\alpha\beta}^{qP} \right )~~~~~~~~\tilde{\varepsilon}_{\alpha\beta}^{qS} = \frac{1}{2}\left ( C_{\alpha\beta}^{qS} - iD_{\alpha\beta}^{qP} \right )
\label{eq:scalar:ceps:2}
\end{equation}
\item[b)] Pseudoescalar parameters:
\begin{equation}
\varepsilon_{\alpha\beta}^{qP} = -\frac{1}{2}\left ( C_{\alpha\beta}^{qP} + iD_{\alpha\beta}^{qS} \right )~~~~~~~~\tilde{\varepsilon}_{\alpha\beta}^{qP} = \frac{1}{2}\left ( C_{\alpha\beta}^{qP} - iD_{\alpha\beta}^{qS} \right )
\label{eq:scalar:ceps}
\end{equation}
\item[c)] Vector parameters:
\begin{equation}
\varepsilon_{\alpha\beta}^{qV} = \frac{1}{2}\left ( C_{\alpha\beta}^{qV} - D_{\alpha\beta}^{qA} \right )~~~~~~~~\tilde{\varepsilon}_{\alpha\beta}^{qV} = \frac{1}{2}\left ( C_{\alpha\beta}^{qV} + D_{\alpha\beta}^{qA} \right )
\label{eq:vector:ceps}
\end{equation}
\item[d)] Tensor parameters: 
\begin{equation}
\varepsilon_{\alpha\beta}^{qT} = \frac{1}{4}\left ( C_{\alpha\beta}^{qT} - iD_{\alpha\beta}^{qT} \right )~~~~~~~~\tilde{ \varepsilon}_{\alpha\beta}^{qT} = \frac{1}{4}\left ( C_{\alpha\beta}^{qT} + iD_{\alpha\beta}^{qT} \right )
\label{eq:tensor:ceps}
\end{equation}
\end{itemize}
In the following sections, we will study the complementarity between different experiments to constrain GNI parameters via neutrino scattering with nucleons and nuclei.  Previous analyses have focused on only one kind of experiment and used different parametrizations. Our goal here is to unify the two parametrizations to explicitly see the interplay between different experiments to constrain GNI.

\section{Neutrino experiments and GNI cross sections }
\label{sec:experiments}
In this section, we review the different neutrino experiments that can be used to constrain GNI parameters. We will show explicitely the corresponding cross-section formulas in terms of a parametrization that can be used on equal footing to study both Coherent Elastic Neutrino-Nucleus Scattering (CEvNS) and Deep Inelastic Scattering (DIS). 
We begin with low-energy experiments measuring the interaction of neutrinos with nuclei, then we move to the high-energy regime, where we analyze neutrino-nucleon interactions in deep-inelastic scattering.

\subsection{Coherent Elastic Neutrino-Nucleus Scattering}
\label{sub:1}
At low energy ranges, neutrinos can interact with an entire nucleus through the neutral current process of CEvNS. Given the coherent nature of the interaction, the predicted SM cross section for this process benefits from an enhancement proportional to $N^2$~\cite{PhysRevD.9.1389},  where $N$ is the number of 
neutrons in the scattering material.  Hence, different sources and detector technologies have been used to measure this interaction. For instance, the COHERENT collaboration has used cesium iodide~\cite{Akimov:2017ade,  COHERENT:2021xmm}, liquid argon~\cite{Akimov:2020pdx} and germanium~\cite{PhysRevLett.134.231801} detectors to measure neutrinos from the decay of pions and muons at rest, while the CONUS+~\cite{Ackermann:2025obx} and Dresden~\cite{COHERENT:2024axu} experiments have used germanium with reactor neutrinos as a source, and the collaborations PANDAX~\cite{PandaX:2024muv}, XENONnT~\cite{XENON:2024ijk}, and LUX-ZEPLIN~\cite{LZ:2025igz} have used xenon to measure signals from solar neutrinos. 

Before studying NSI within the context of CEvNS interactions, notice that the Lagrangians in Eqs.~\eqref{eq:GNI} and~\eqref{eq:GNI:D} contain field operators defined at a quark level. However, we need to define operators at a nuclear level for CEvNS. The transition from neutrino-quark to neutrino-nucleus operators has been studied within the context of the `\textit{C}' parametrization in \cite{Lindner:2016wff, AristizabalSierra:2018eqm, Bischer:2019ttk}, which can be translated into the `\textit{epsilon}' parametrization through Eqs.~\eqref{eq:scalar:ceps:2} to \eqref{eq:tensor:ceps}. For the particular case of scalar interactions, the contribution to the cross section is~\cite{Lindner:2016wff}:

\begin{eqnarray}
\label{eq:cevns:scalar:c}
  \frac{\mathrm{d}\sigma_S}{\mathrm{d}T} \, =\ \frac{G_F^2 M}{\pi}\ F^2(|\vec{q}|^2)
  && \left[\,  \left | \sum_{q = u,d} \left( Z\, \frac{m_p}{m_q}f_q^p + N\, \frac{m_n}{m_q}f_q^n \right)
C_{\alpha\beta}^{qS} \right |^2   \right.  \nonumber  \\
  & & \left.  + \left | \sum_{q = u,d} \left( Z\, \frac{m_p}{m_q}f_q^p + N\, \frac{m_n}{m_q}f_q^n \right)
D_{\alpha\beta}^{qP} \right |^2 \right]\, \frac{MT}{8E_\nu^2} ,
\end{eqnarray} 
which in the `\textit{epsilon}' parametrization has the form
\begin{eqnarray}
\label{eq:cevns:scalar}
  \frac{\mathrm{d}\sigma_S}{\mathrm{d}T} \, =\ \frac{G_F^2 M}{\pi}\ F^2(|\vec{q}|^2)
  && \left[\,  \left | \sum_{q = u,d} \left( Z\, \frac{m_p}{m_q}f_q^p + N\, \frac{m_n}{m_q}f_q^n \right)
\left ( \varepsilon_{\alpha\beta}^{qS} +\tilde{\varepsilon}_{\alpha\beta}^{qS} \right ) \right |^2   \right.  \nonumber  \\
  & & \left.  + \left | \sum_{q = u,d} \left( Z\, \frac{m_p}{m_q}f_q^p + N\, \frac{m_n}{m_q}f_q^n \right)
\left ( \varepsilon_{\alpha\beta}^{qS} -\tilde{\varepsilon}_{\alpha\beta}^{qS} \right ) \right |^2 \right]\, \frac{MT}{8E_\nu^2} ,
\end{eqnarray} 
where $m_p$ and $m_n$ are the mass of the proton and neutron, respectively, and the coupling constants $f_q^N$ are given in Table \ref{tab:ct}.  On the other hand, $M$ is the mass of the target material, $T$ is the nuclear recoil energy that results from the interaction,  and $E_\nu$ is the energy of the incoming neutrino. Notice that Eqs.~\eqref{eq:cevns:scalar:c} and~\eqref{eq:cevns:scalar} also depend on the form factor $F(q^2)$, which is introduced to parametrize the distributions of protons and neutrons within the nucleus. In general, the most widely accepted form  of this function is the Klein-Nystrand parametrization, which we use for our computations. However, the use of other parametrizations, such as the Helm or Symmetrized Fermi, does not drastically change the results.

In the case of vector interactions, the CEvNS cross section in the `\textit{C}' parametrization takes the form
\begin{eqnarray}
  \frac{\mathrm{d}\sigma_V}{\mathrm{d}T} (E_\nu, T)\, =\ \frac{G_F^2 M}{\pi}\ F^2(|\vec{q}|^2)\!
  \left[\,  \left|  \frac{1}{2}\left(C_{\alpha\beta}^{V} - D_{\alpha\beta}^{A}\right)+ Q_{W, \alpha}\, \delta_{\alpha \beta} \right|^2 \left ( 1-\frac{MT}{2E_\nu^2} \right )    \right]\, ,
\label{eq:CEvNS_GNI}
\end{eqnarray}
where $Q_{W,\alpha} = g_V^pZ + g_V^nN$ is the weak charge with predicted values within the SM of $g_V^p = 1/2 - 2\sin^2\theta_W$, and $g_V^n = -1/2$, being $\theta_W$ the weak mixing angle.  In the previous equation we have used the definitions
\begin{equation}
C^V_{\alpha \beta}\, =\, (2 Z + N)\, C^{uV}_{\alpha \beta} + (Z + 2 N)\, C^{dV}_{\alpha \beta}\ ,
\label{eq:vectorC}
\end{equation}
and
\begin{equation}
D^A_{\alpha \beta}\, =\, (2 Z + N)\, D^{uA}_{\alpha \beta} + (Z + 2 N)\, D^{dA}_{\alpha \beta}\ .
\label{eq:vectorD}
\end{equation}
 For an incident antineutrino of flavor $\alpha$, one should replace $C^S_{\beta \alpha} \to C^S_{\alpha \beta}$,
$C^V_{\beta \alpha} \to C^V_{\alpha \beta}$ and $C^T_{\beta \alpha} \to C^T_{\alpha \beta}$
in Eq.~(\ref{eq:CEvNS_GNI}) \cite{Chatterjee:2024vkd}. 
As in the scalar case, we can rewrite Eq.~\eqref{eq:CEvNS_GNI} in the `\textit{epsilon}' notation. 
In the absence of right handed neutrinos, by using Eq. \eqref{eq:vector:ceps}, we recover the well known expression~\cite{Barranco:2005yy}

\begin{eqnarray}
  \frac{\mathrm{d}\sigma_V}{\mathrm{d}T} \, =\ \frac{G_F^2 M}{\pi}\ F^2(|\vec{q}|^2)\!
   \left[\, \left|  (2Z+N)\varepsilon_{\alpha\beta}^{uV} + (Z+2N)\varepsilon_{\alpha\beta}^{dV}   + Q_{W, \alpha}\, \delta_{\alpha \beta} \right|^2 \left( 1 - \frac{MT}{2E_\nu^2} \right)  \right]\, .
\label{cross.vector}
\end{eqnarray} 

Moving now to the case of tensor interactions, it has been recently shown that, contrary to the scalar and vector cases, tensor interactions are not coherently enhanced in CEvNS \cite{Breso-Pla:2025pds}. However,  its associated cross section is not entirely suppressed by nuclear spin, taking the form~\cite{Liao:2025hcs}:

\begin{eqnarray}
\label{eq:cevns:tensor}
 \frac{\mathrm{d}\sigma_T}{\mathrm{d}T} \, =\ \frac{8G_F^2 M^2T}{\pi M_A^2}\ 
  && \sum_{a,a' = p,n}  \left[\,   \left( \delta_T^{ua}\varepsilon_{\alpha\beta}^{uT} + \delta_T^{da}\varepsilon_{\alpha\beta}^{dT} \right)
\left ( \delta_T^{ua'}\varepsilon_{\alpha\beta}^{uT} + \delta_T^{da'}\varepsilon_{\alpha\beta}^{dT}  \right )\mathcal{F}_a^M \mathcal{F}_{a'}^M    \right.  \nonumber  \\
  && \left.  + 4\left( \delta_T^{ua}\varepsilon_{\alpha\beta}^{uT} + \delta_T^{da}\varepsilon_{\alpha\beta}^{dT} \right)
\left ( F_{1T}^{ua'}\varepsilon_{\alpha\beta}^{uT} + F_{1T}^{da'}\varepsilon_{\alpha\beta}^{dT}  \right )\mathcal{F}_a^{\Phi ''} \mathcal{F}_{a'}^M    \right.  \nonumber  \\
  & & \left.  + 4 \left ( F_{1T}^{ua}\varepsilon_{\alpha\beta}^{uT} + F_{1T}^{da}\varepsilon_{\alpha\beta}^{dT}  \right )\left ( F_{1T}^{ua'}\varepsilon_{\alpha\beta}^{uT} + F_{1T}^{da'}\varepsilon_{\alpha\beta}^{dT}  \right )\mathcal{F}_a^{\Phi ''} \mathcal{F}_{a'}^{\Phi ''}  \right]\, \left(1- \frac{T}{E_\nu} \right) ,
\end{eqnarray} 

\noindent where we have used the notation $\delta_{T}^{qn} = 2F_{2T}^{qn}-F_{1T}^{qn}$, and the values of $F_{iT}^{qn}$ are given in Table \ref{tab:ct}.  On the other hand, the factors $F_{a}^{M}$ and $F_{a}^{\Phi''}$ depend on the target material and were extracted from Ref. \cite{Hoferichter:2020osn}.

\begin{table}[]
\centering
\resizebox{\textwidth}{!}{%
\begin{tabular}{|c|c|cc|}
\hline
\textbf{Coupling}        & \textbf{Scalar}             & \multicolumn{2}{c|}{\textbf{Tensor}}                                                     \\ \hline
\multirow{2}{*}{$q = u$} & $~~~~f_u^p = 0.0197 \pm 0.0014~~~~$ & \multicolumn{1}{c|}{$~~~~F^{up}_{1T} = 0.784 \pm 0.028~~~~$}  & $~~~~F^{un}_{1T} = -0.204 \pm 0.014~~~~$ \\ \cline{2-4} 
                         & $f_u^n = 0.0178 \pm 0.0013$ & \multicolumn{1}{c|}{$F^{up}_{2T} = -1.5 \pm 1.0$}     & $F^{un}_{2T} = 0.5 \pm 0.3$      \\ \hline
\multirow{2}{*}{$q = d$} & $f_d^p = 0.0383 \pm 0.0027$ & \multicolumn{1}{c|}{$F^{dp}_{1T} = -0.204 \pm 0.014$} & $F^{dn}_{1T} = 0.784 \pm 0.028$  \\ \cline{2-4} 
                         & $f_d^n = 0.0423 \pm 0.0026$ & \multicolumn{1}{c|}{$F^{dp}_{2T} = 0.5 \pm 0.3$}      & $F^{dn}_{2T} = -1.5 \pm 1.0$     \\ \hline
\end{tabular}%
}
\caption{Scalar~\cite{Hoferichter:2023ptl} and tensor~\cite{Gupta:2018lvp} couplings used for the generalized cross sections defined in Eqs.  \eqref{eq:cevns:scalar} and \eqref{eq:cevns:tensor}, respectively.}
\label{tab:ct}
\end{table}

We can use COHERENT data to constrain GNI parameters individually and compare with constraints from other processes. For the latest CsI measurement\footnote{As shown in Ref. \cite{DeRomeri:2022twg}, the sensitivity from CsI data is dominant over LAr data. Hence, here we only include the analysis from CsI. }, the experimental number of events, $N_{ij}^{exp}$, has been provided in $i = 9$ recoil energy bins and $j = 11$ arrival timing bins\footnote{The number of events for each bin is given in the appendix of Ref. \cite{DeRomeri:2022twg}.}.  Then, to set constraints on GNI parameters, we compare the predicted number of events, $N_{ij}^{th}$, with the experimental counterpart by minimizing the squared function 
\begin{equation}
  \chi^2 = 2\sum_{i,j}\left [ N^{\textrm{th}}_{ij}(\varepsilon_{\alpha\beta}^{qX}) - N^{\textrm{exp}}_{ij} + N^{\textrm{exp}}_{ij}\ln\left ( \frac{N^{\textrm{exp}}_{ij}}{N^{\textrm{th}}_{ij}(\varepsilon_{\alpha\beta}^{qX})} \right )\right ] + \sum_{m=0}^2 \frac{\alpha_{m}}{\sigma_{\alpha_m}^2} + \sum_{k=1}^3\frac{\beta_k}{\sigma_{\beta_k}^2} ,
    \label{eq:chi:CsI}
\end{equation}
where we have
\begin{equation}
    N^{\textrm{th}}_{ij}(\varepsilon_{\alpha\beta}^{qX}) = (1+\alpha_0) N^{\textrm{CE$\nu$NS}}_{ij}(\varepsilon_{\alpha\beta}^{qX},\alpha_1, \alpha_2,\alpha_3) +  (1+\beta_1)N^{\textrm{SSB}}_{ij}+  (1+\beta_2)N^{\textrm{BRN}}_{ij}(\alpha_3)+  (1+\beta_3)N^{\textrm{NIN}}_{ij}(\alpha_3) \,.
    \label{N:chi:CsI}
\end{equation}
Notice that in the last equation, the expected number of events is obtained by considering the predicted contributions from CEvNS, as well as three main backgrounds sources. These include  steady state background (SSB),  beam-related neutrons (BRN), and neutrino-induced neutrons (NIN). The CEvNS contribution is computed as a convolution of the cross section, which depends on the different GNI,  with the total neutrino flux,  considering the efficiency, signal resolution, and timing distribution as given in the supplemental material of Ref.  \cite{COHERENT:2021xmm}.  By following the procedure given in Ref. \cite{DeRomeri:2022twg}, we have included a total of 7 nuisance parameters in Eqs.~(\ref{eq:chi:CsI}) and (\ref{N:chi:CsI}), with respect to which the $\chi^2$ function is minimized, each with its corresponding uncertainty $\sigma_{\alpha_i}$, $\sigma_{\beta_j}$\footnote{The values of $\sigma_{\alpha_i}$, $\sigma_{\beta_j}$ are taken from Ref. \cite{DeRomeri:2022twg}}.  As described on the same reference, the parameters $\alpha_0$, $\alpha_1$, and $\alpha_2$, are related to CEvNS signal normalization, form factor, and efficiency, respectively, while $\beta_1$, $\beta_2$, and $\beta_3$ are related to normalization uncertainties of SSB, BRN, and NIN backgrounds, respectively, and $\alpha_3$ is related to the timing efficiency, which is common for CEvNS signal, BRN, and NIN.

\subsection{Neutrino-nucleon Deep-Inelastic  Scattering}

High-energy neutrinos (tens of GeV to a few TeV) interact with quarks inside nucleons via charged- and neutral-current weak interactions, leading to the breakup of the nucleon and the production of a hadronic final state. 
The CHARM~\cite{Dorenbosch:1986tb,Allaby:1987vr} and CDHS~\cite{Blondel:1989ev} experiments, both located at CERN, played a central role in the study of neutrino–nucleon deep inelastic scattering~\footnote{The NuTeV experiment at Fermilab also measured DIS with great precision, but since the analysis is very sensitive to many hadronic and nuclear uncertainties, we did not include it in this work.}. Operating with high-energy neutrino beams in the range of 20–300 GeV, these experiments provided some of the most precise measurements of neutrino cross sections in the DIS regime.
In addition to its role in studying quark dynamics and measuring the weak mixing angle, DIS also provides sensitivity to possible deviations from the SM in the form of generalized neutrino interactions. 

The total neutral-current cross section for neutrinos and antineutrinos has the following expressions, respectively:
\begin{equation}
	\begin{split}
		\sigma^\mathrm{NC}_{\nu N,\mathrm{SM}}=&\frac{G^{2}_{F}s}{2\pi}\Big[\left((g^{uL})^{2}+\frac{1}{3}(g^{uR})^{2}\right)f_{q}+\left((g^{dL})^{2}+\frac{1}{3}(g^{dR})^{2}\right)f_{q}\\
		&+\left((g^{uR})^{2}+\frac{1}{3}(g^{uL})^{2}\right)f_{\bar{q}}+\left((g^{dR})^{2}+\frac{1}{3}(g^{dL})^{2}\right)f_{\bar{q}}\Big]\,,
	\end{split}
	\label{eq:nuQuarkCrossSec_2}
\end{equation}
\begin{equation}
	\begin{split}
		\sigma^\mathrm{NC}_{\bar{\nu}N,\mathrm{SM}}=&\frac{G^{2}_{F}s}{2\pi}\Big[\left((g^{uR})^{2}+\frac{1}{3}(g^{uL})^{2}\right)f_{q}+\left((g^{dR})^{2}+\frac{1}{3}(g^{dL})^{2}\right)f_{q}\\
		&+\left((g^{uL})^{2}+\frac{1}{3}(g^{uR})^{2}\right)f_{\bar{q}}+\left((g^{dL})^{2}+\frac{1}{3}(g^{dR})^{2}\right)f_{\bar{q}}\Big]\,,
	\end{split}
	\label{eq:nuQuarkCrossSec_2p}
\end{equation}
where we take the following values for the SM effective couplings~\cite{Erler:2013xha}:
\begin{equation}
	g^{uL} = 0.3457 , \quad g^{uR} = -0.1553 , \quad g^{dL} = -0.4288, \quad g^{dR} = 0.0777.
\end{equation}
Assuming an isoscalar target, $f_q$ ($f_{\bar{q}}$) are nuclear parton distribution functions accounting for the fraction of proton momentum corresponding to up and down quarks (antiquarks)~\cite{Martin:2009iq}. Including GNI effects from the Lagrangian in Eq.~\eqref{eq:epsilon}, the total cross sections for scalar (pseudoscalar) and tensor interactions are given by~\cite{Han:2020pff} 
\begin{equation}
	\sigma^\mathrm{NC}_{\nu N,S(P)}=\sigma^\mathrm{NC}_{\bar{\nu} N,S(P)}=\frac{G^{2}_{F}s}{24\pi}\sum_{q=u,d}\sum_{\beta} \left[|\eps^{qS(P)}_{\alpha\beta} + \tilde{\eps}^{qS(P)}_{\alpha\beta}|^{2}\left(\frac{f_{q}+f_{\bar{q}}}{2}\right)\right]\,,
	\label{eq:nuQuarkCrossSec_3}
\end{equation} 
\begin{equation}
	\sigma^\mathrm{NC}_{\nu N,T}=\sigma^\mathrm{NC}_{\bar{\nu} N,T}=\frac{28G^{2}_{F}s}{3\pi}\sum_{q=u,d}\sum_{\beta} \left[|\eps^{qT}_{\alpha\beta} + \tilde{\eps}^{qT}_{\alpha\beta}|^{2}\left(\frac{f_{q}+f_{\bar{q}}}{2}\right)\right]\, .
	\label{eq:nuQuarkCrossSec_4}
\end{equation} 

The cross section for new vector interactions takes the same form as in Eqs.~\eqref{eq:nuQuarkCrossSec_2} and~\eqref{eq:nuQuarkCrossSec_2p}, with the following changes
\begin{equation}
	(g^{qL})^2 \to (\tilde{g}^{qL})^2 = (g^{qL} + \eps^{qL}_{\alpha\alpha} )^2 + \sum_{\beta\neq\alpha}|\eps^{qL}_{\alpha\beta}|^2 ,
	\label{eq:gLtilde}
\end{equation}
\begin{equation}
	(g^{qR})^2 \to (\tilde{g}^{qR})^2 = (g^{qR} + \eps^{qR}_{\alpha\alpha} )^2 + \sum_{\beta\neq\alpha}|\eps^{qR}_{\alpha\beta}|^2 ,
	\label{eq:gRtilde}
\end{equation}
where $\eps^{qL}_{\alpha\beta} = \frac{1}{2}(\eps^{qV}_{\alpha\beta} + \eps^{qA}_{\alpha\beta})$ and $\eps^{qR}_{\alpha\beta} = \frac{1}{2}(\eps^{qV}_{\alpha\beta} - \eps^{qA}_{\alpha\beta})$.
The CHARM collaboration measured the ratio of total neutral- to charged-current cross-section for semileptonic 
scattering by using an electron-neutrino beam with equal $\nu_e$ and $\bar{\nu}_e$ fluxes. This ratio is defined as~\cite{Dorenbosch:1986tb,Allaby:1987vr}
\begin{equation}
	R^e \equiv \frac{\sigma(\nu_e N \to \nu X) + \sigma(\bar{\nu}_e N \to \bar{\nu} X)}{\sigma(\nu_e N \to e^- X) + \sigma(\bar{\nu}_e N \to e^+ X)}, 
\end{equation}
with a measurement of $R^e_\mathrm{exp} = 0.406 \pm 0.140$~\cite{Dorenbosch:1986tb}, while the SM prediction
takes the value $R^e_\mathrm{SM} = g_{L}^2 + g_{R}^2 = 0.335$. When including new interactions, this ratio takes the form~\cite{Han:2020pff} 
\begin{equation}
	R^{e}_\mathrm{th} = \tilde{g}_{L}^2 + \tilde{g}_{R}^2 + \frac{1}{12}\sum_{q=u,d}\sum_{\beta}\left(|\eps^{qS}_{e\beta} + \tilde{\eps}^{qS}_{e\beta}|^2 + |\eps^{qP}_{e\beta} + \tilde{\eps}^{qP}_{e\beta}|^2 + 224 |\eps^{qT}_{e\beta} + \tilde{\eps}^{qT}_{e\beta}|^2 \right),
\end{equation}
where $\tilde{g}_{L}^2 = (\tilde{g}^{uL})^2 + (\tilde{g}^{dL})^2$ and $\tilde{g}_{R}^2 = (\tilde{g}^{uR})^2 + (\tilde{g}^{dR})^2$,  according with the definitions of Eqs.~\eqref{eq:gLtilde} and~\eqref{eq:gRtilde}. Notice that for this case, the first flavor index of the GNI parameters is set to $\alpha=e$.

The following $\chi^2$ function is defined to perform the statistical analysis on the GNI parameters:
\begin{equation}
	\chi^2_{\mathrm{CHARM}-e} = \left(\frac{R^{e}_\mathrm{th} - R^e_\mathrm{exp}}{\sigma^e}\right)^2.
\end{equation}

Additionally,  the experiments CDHS and CHARM measured the ratios
of neutral-to-charged-current semileptonic cross-sections, $R^\nu$ and $R^{\bar{\nu}}$,
using a muon-neutrino beam. These ratios, defined as
\begin{equation}
	R^\nu \equiv \frac{\sigma(\nu_\mu N \to \nu X)}{\sigma(\nu N \to \mu^- X)}, \quad 	R^{\bar{\nu}} \equiv \frac{\sigma(\bar{\nu}_\mu N \to \bar{\nu} X)}{\sigma(\bar{\nu} N \to \mu^+ X)},
\end{equation}
take the following form in the SM framework
\begin{equation}
	R^\nu_\mathrm{SM} = g_L^2 + r g_R^2, \quad R^{\bar{\nu}}_\mathrm{SM} = g_L^2 + \frac{1}{r} g_R^2,
\end{equation}
where the parameter $r$ corresponds to the ratio of charged-current cross-sections~\cite{Jonker:1980vf}
\begin{equation}
	r = \frac{\sigma(\bar{\nu} N \to \mu^+ X)}{\sigma(\nu N \to \mu^- X)}
	\label{eq:parameter_r} = \frac{\tfrac{1}{3}f_q + f_{\bar{q}}}{f_q + \tfrac{1}{3}f_{\bar{q}}} .
\end{equation} 
The experimental values of $R^\nu_\mathrm{exp}$, $R^{\bar{\nu}}_\mathrm{exp}$, and $r$ measured by CHARM and CDHS are presented in Table~\ref{Tab:DIS_measurements}.

\begin{table}[ht]
	\centering
	\begin{tabular}{c@{\hskip 0.2in}c@{\hskip 0.2in}c@{\hskip 0.2in}c} \hline
		\toprule
		& $R^\nu_\mathrm{exp}$ & $R^{\bar{\nu}}_\mathrm{exp}$ & $r$  \\   \hline  \hline
		CHARM~\cite{Allaby:1987vr} & $0.3093 \pm 0.0031$ & $0.390 \pm 0.014$ & $0.456 \pm 0.011$ \\
		CDHS~\cite{Blondel:1989ev} & $0.3072 \pm 0.0033$ & $0.382 \pm 0.016$ & $0.393 \pm 0.014$  \\  \hline
	\end{tabular} 
	\caption{\footnotesize Experimental values of the neutral to charged-current cross 
		section ratios $R^\nu$ and 
		$R^{\bar{\nu}}$, and the ratio $r$ defined in Eq.~\eqref{eq:parameter_r}, reported by the CHARM and CDHS collaborations .
	} 
	\label{Tab:DIS_measurements}    
\end{table}

Using Eqs.~\eqref{eq:nuQuarkCrossSec_3} and
\eqref{eq:nuQuarkCrossSec_4}, we can compute the ratios expressions when including GNI~\cite{Han:2020pff}
\begin{eqnarray}
	R^\nu_\mathrm{th} = \tilde{g}_L^2 + r \tilde{g}_R^2 + \frac{1}{32}(1+r)\sum_{q=u,d}\sum_\beta \left(|\eps^{qS}_{\mu\beta} + \tilde{\eps}^{qS}_{\mu\beta}|^2 + |\eps^{qP}_{\mu\beta}+\tilde{\eps}^{qP}_{\mu\beta}|^2 + 224 |\eps^{qT}_{\mu\beta} + \tilde{\eps}^{qT}_{\mu\beta}|^2 \right), \nonumber \\
	R^{\bar{\nu}}_\mathrm{th} = \tilde{g}_L^2 + \frac{1}{r} \tilde{g}_R^2 +  \frac{1}{32}(1+\frac{1}{r})\sum_{q=u,d}\sum_\beta \left(|\eps^{qS}_{\mu\beta} + \tilde{\eps}^{qS}_{\mu\beta}|^2 + |\eps^{qP}_{\mu\beta}+\tilde{\eps}^{qP}_{\mu\beta}|^2 + 224 |\eps^{qT}_{\mu\beta} + \tilde{\eps}^{qT}_{\mu\beta}|^2 \right).
	\label{eq:ratiosGNI_CHARM_CDHS}
\end{eqnarray}

Notice that, since we only study the effect of GNI in neutral-current interactions, the ratio $r$ is unaffected by the presence of new interactions.

To analyze the data from CHARM and CDHS, we have adopted the following $\chi^2$ function that includes the correlation between the neutrino and antineutrino ratios
\begin{equation}
	\chi^2_{\mathrm X} = \sum_{j,k=\nu,\bar{\nu}}(R^j_\mathrm{th} - R^j_{\mathrm{exp}}
                )(\sigma^2)^{-1}_{jk}(R^k_\mathrm{th} - R^k_{\mathrm{exp}}
                ),
\end{equation}
where $\sigma^2$ is the covariance matrix constructed from the squared
uncertainties, $R^j_\mathrm{th}$ is the neutrino or antineutrino
ratio including GNI, $R^j_\mathrm{exp}$
is the experimental value, and $\mathrm{X}$ stands for either $\mathrm{CHARM}$ or $\mathrm{CDHS}$.

The main benefit of mapping between the two GNI parametrizations introduced in Section~\ref{sec:theory} is to set constraints from experiments focused on different energy regimes on the same footing.  In practice, low-energy CEvNS data and high-energy deep-inelastic scattering results are usually described with different operator bases, making it difficult to directly compare their sensitivities to new physics under the GNI formalism. By rewriting all observables in a single parametrization, we can directly compare the experimental bounds from COHERENT and CHARM/CDHS,  visualizing the type of interaction for which each experiment is more sensitive,  and making their complementarity explicit.

\section{Constraints}
\label{sec:results}
\begin{figure}[t]
\centering
\includegraphics[scale=0.7]{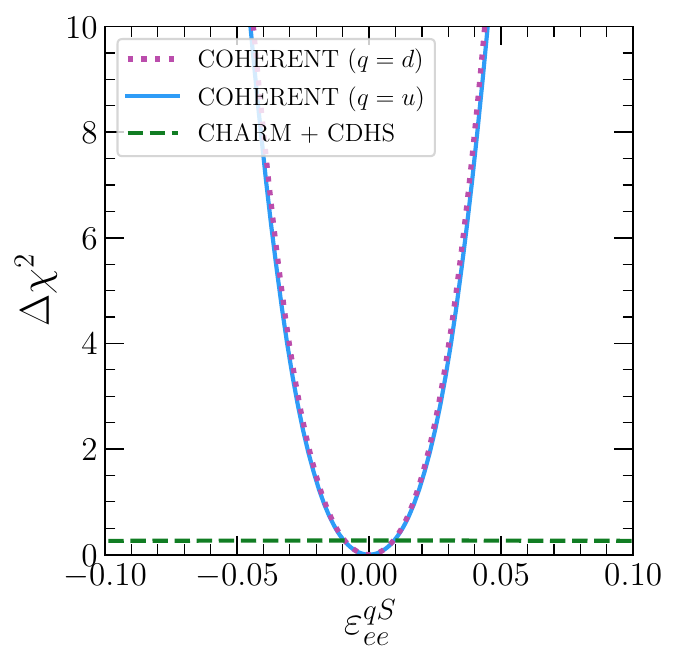}
\includegraphics[scale=0.7]{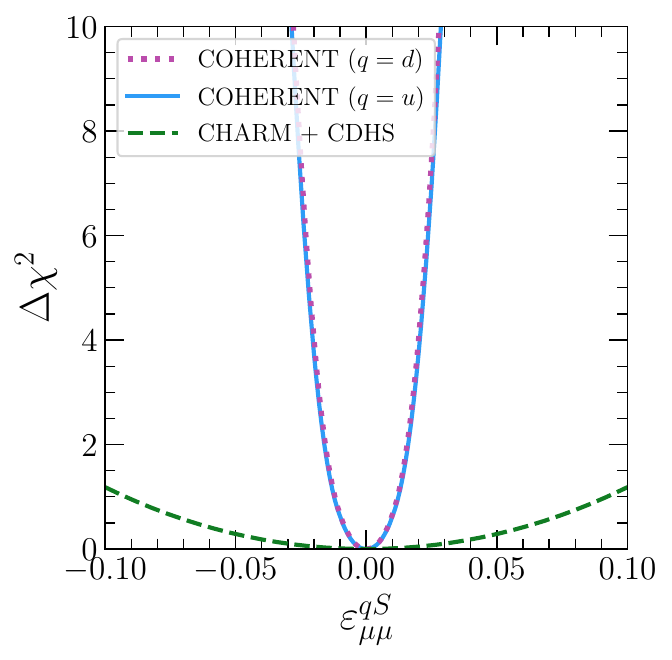}
\includegraphics[scale=0.7]{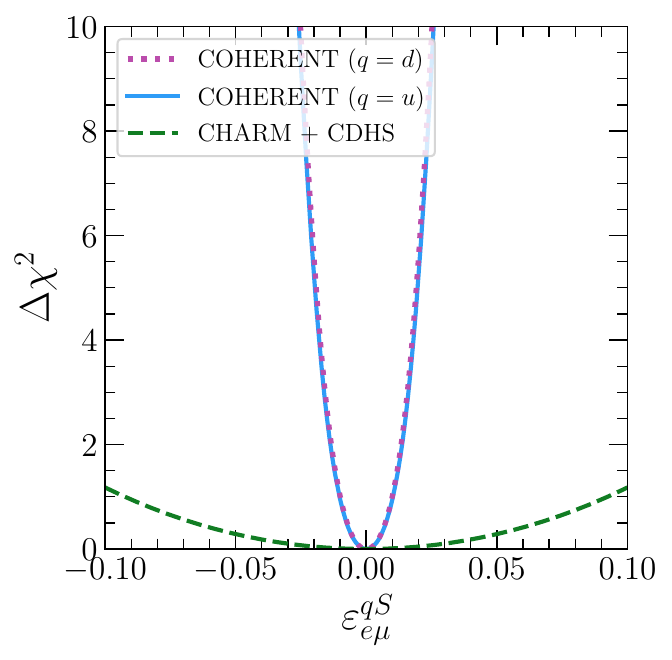}
\caption{$\Delta\chi^2$ profile for scalar interactions of up and down quark couplings obtained from COHERENT (solid blue and dotted magenta) and CHARM + CDHS (dashed green) data. Top panels correspond to $\varepsilon_{ee}^{qS}$ and $\varepsilon_{\mu\mu}^{qS}$, while the bottom panel corresponds to $\varepsilon_{e\mu}^{qS}$.}
\label{Result-Scalar} 
\end{figure}

We now provide a summary of the constraints that the different neutrino experiments described in the previous section can give on individual GNI parameters.  We begin by showing in Fig.~\ref{Result-Scalar} the results for scalar parameters, assuming only one of them to be different from zero at a time and assuming no interaction of either tensor or vector type present in the analysis.  Notice that, given the shape of the scalar contribution cross section in Eq.~(\ref{eq:cevns:scalar}), the obtained constraints for $\varepsilon_{\alpha\beta}^{qS}$ are the same as those for $\tilde{\varepsilon}_{\alpha\beta}^{qS}$. Hence, we only show the results for $\varepsilon_{\alpha\beta}^{qS}$. 
The top left panel in the figure shows the results for $\varepsilon_{ee}^{qS}$.  In this case, the COHERENT experiment shows better sensitivity, with the solid blue line corresponding to the case of an \textit{up}-quark coupling and the dotted violet line to the \textit{down}-quark  coupling. We notice that in both cases, the sensitivity is almost the same, while the combination CHARM+CDHS, shown in dashed green, has poor sensitivity.
A similar analysis is shown in the top-right panel of the same figure for the parameter $\varepsilon_{\mu\mu}^{qS}$, using the same color code as in the previous case.  Here, we notice again that the COHERENT analysis results in a better sensitivity. 
Finally, we show in the bottom panel of Fig.~\ref{Result-Scalar} the results for the flavor changing parameter $\varepsilon_{e\mu}^{qS}$, following a similar behaviour as those from the previous two cases. We conclude that, for scalar interactions, COHERENT bounds are better than those from deep inelastic scattering by at least one  order of magnitude, with the following results at a 90\% C.L.:
\begin{equation}
\begin{array}{rcll}
-0.026 \;<\; \varepsilon_{ee}^{qS} \;<\; 0.026 
  &&& \text{(COHERENT)}
  \\[10pt]

-0.017 \;<\; \varepsilon_{\mu\mu}^{qS} \;<\; 0.017
  &&& \text{(COHERENT)}
  \\[10pt]

-0.015 \;<\; \varepsilon_{ee}^{qS} \;<\; 0.015
  &&& \text{(COHERENT)}
\end{array}
\end{equation}
\begin{figure}
\centering
\includegraphics[scale=0.62]{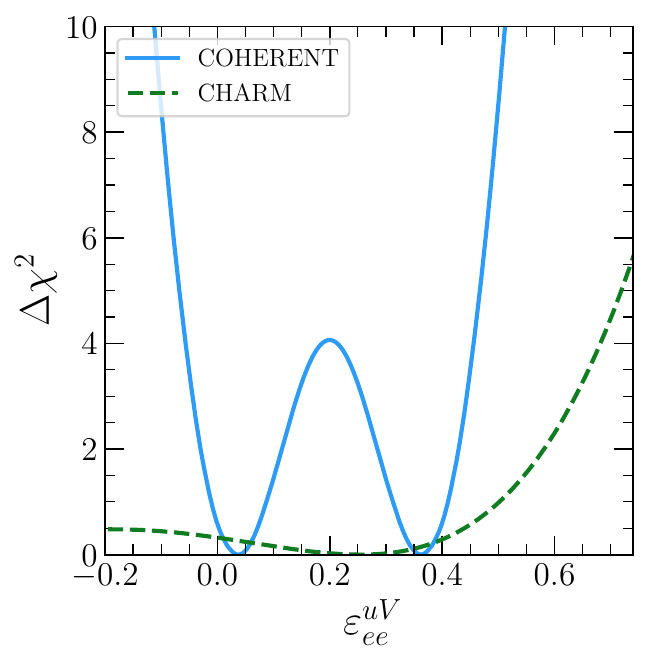}
\includegraphics[scale=0.62]{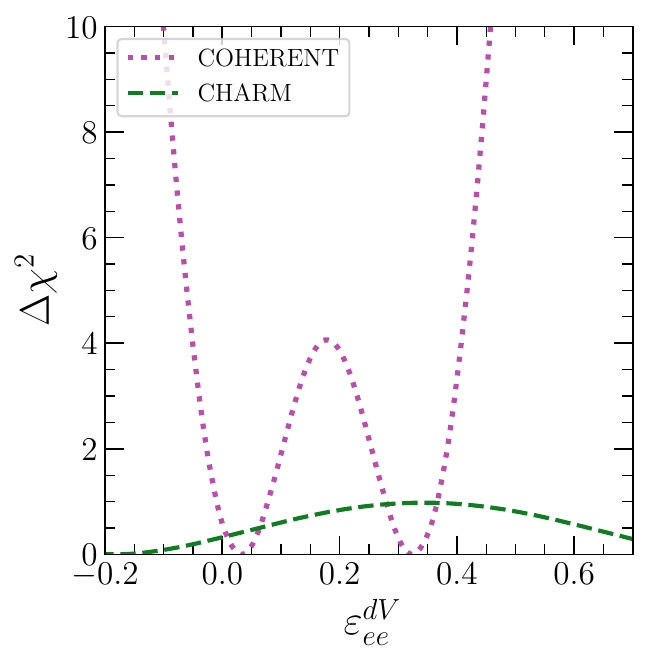}
\vskip .4cm
\includegraphics[scale=0.62]{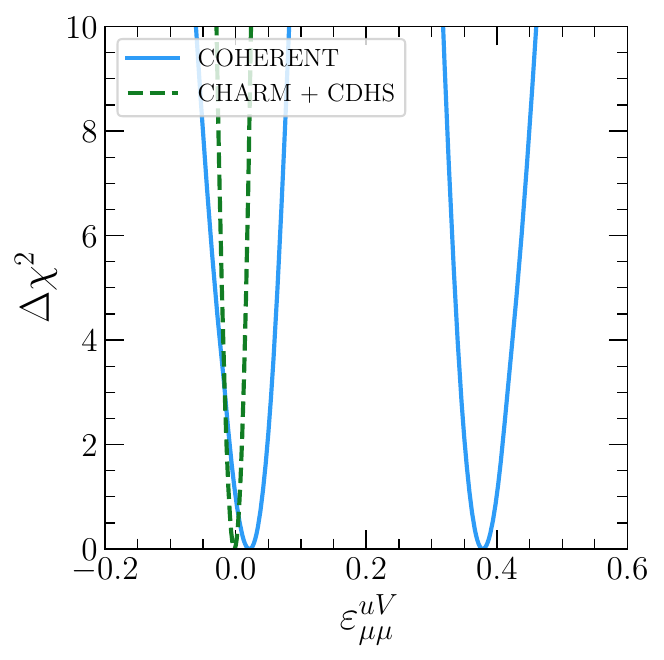}
\includegraphics[scale=0.62]{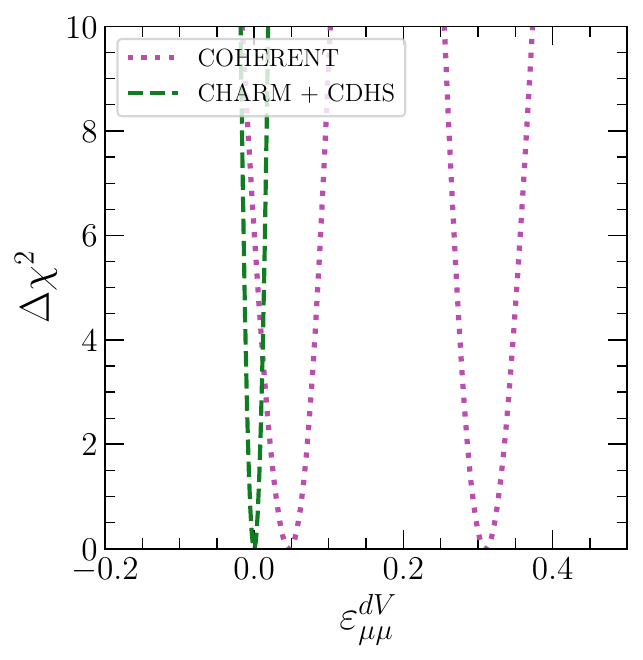}
\caption{$\Delta\chi^2$ profile for non-universal vector interactions of up and down quark couplings obtained from COHERENT (solid blue and dotted magenta) and CHARM + CDHS (dashed green) data. Top panels correspond to electron neutrino
parameters, while the bottom panels corresponds to muon neutrino parameters.}
\label{Result:Vector:1} 
\end{figure}

We now turn our attention to the case of vector interactions, also known as NSI.  In Fig.~\ref{Result:Vector:1}, we show the results for non-universal parameters, with the top panels corresponding to the case of $\varepsilon_{ee}^{qV}$. On the left side, we show the results for $\varepsilon_{ee}^{uV}$, with the blue and green lines indicating the COHERENT and CHARM bounds, respectively. Notice that the COHERENT profile shows the presence of two minima, in agreement with the shape of the cross section in Eq.~\eqref{cross.vector}. The CHARM profile also has the presence of two minima, but the scale used in the figure displays only one, showing that the COHERENT results are more restrictive in this case.  Similarly, the top-right panel shows the profiles for the parameter $\varepsilon_{ee}^{dV}$, with the COHERENT result in magenta and the CHARM result in dashed green.  Again, we notice the two minima for COHERENT in the displayed range, showing a better sensitivity from this experiment.

The bottom panels in Fig.~\ref{Result:Vector:1} correspond to the non-universal parameters $\varepsilon_{\mu\mu}^{qV}$. The left panel shows the obtained bounds for $\varepsilon_{\mu\mu}^{uV}$.  In contrast to previous results, these panels show that the combination CHARM+CDHS (dashed green) gives better bounds when compared to the COHERENT (blue) result.  This is expected since both CHARM and CDHS primarily produce muon (anti)neutrinos, having better statistics and less systematic effects when compared to the COHERENT experiment.  The right panel in the same figure shows the corresponding result for $\varepsilon_{\mu\mu}^{dV}$. Overall, we
observe a similar behavior as in the previous case, with the CHARM+CDHS (dashed green) profile more restrictive than the COHERENT (magenta) one.  However, given the resolution of CHARM, we notice a slight tension between the two results at 90\% C.L., corresponding to a value of $\Delta\chi^2 = 2.71$. In summary, we obtain the following bounds for non-universal NSI:

\begin{equation}
\begin{array}{rcll}
-0.040 \;<\; \varepsilon_{ee}^{uV} \;<\; 0.133~~ 
  \cup~~ 0.265 \;<\; \varepsilon_{ee}^{uV} \;<\; 0.440
  & \text{(COHERENT)}
  \\[10pt]

-0.034 \;<\; \varepsilon_{ee}^{dV} \;<\; 0.120~~
  \cup~~ 0.235 \;<\; \varepsilon_{ee}^{dV} \;<\; 0.393
  & \text{(COHERENT)}
  \\[10pt]

-0.013 \;<\; \varepsilon_{\mu\mu}^{uV} \;<\; 0.013~~~~~~~~~~~~~~~~
  
  & \text{(CHARM+CDHS)}
  \\[10pt]

-0.010 \;<\; \varepsilon_{\mu\mu}^{dV} \;<\; 0.010~~~~~~~~~~~~~~~~
  
  & \text{(CHARM+CDHS)}
  \\[10pt]
  
\end{array}
\end{equation}
\begin{figure}
\centering
\includegraphics[scale=0.70]{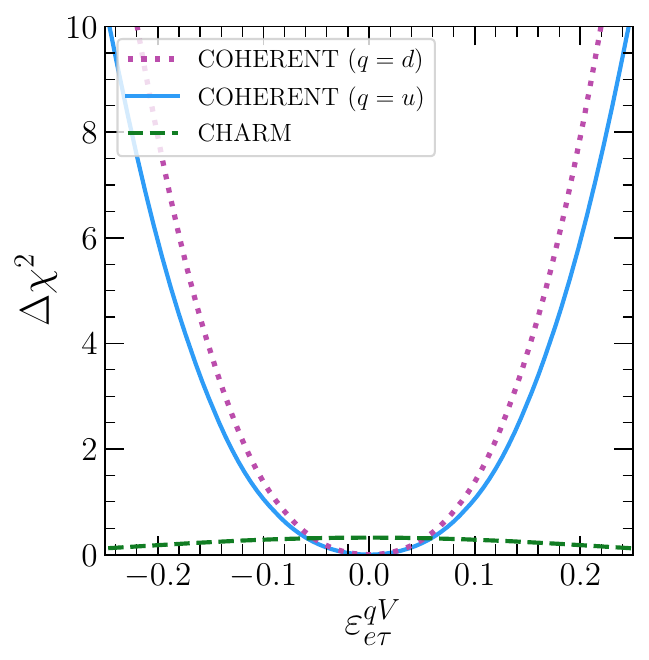}
\includegraphics[scale=0.70]{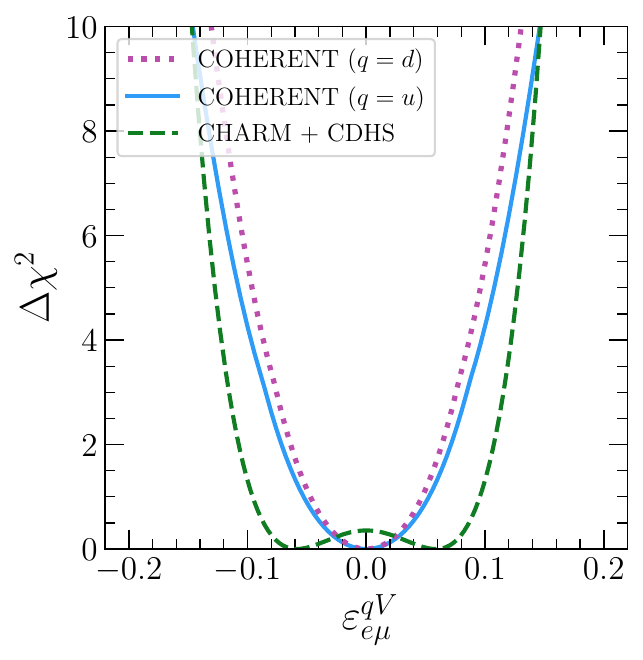}
\vskip .4cm
\includegraphics[scale=0.70]{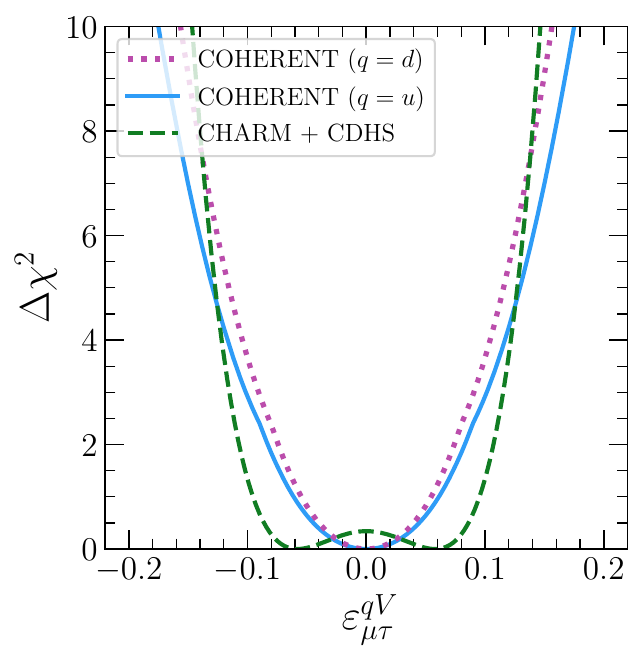}
\caption{$\Delta\chi^2$ profile for flavor changing vector interactions of up and down quark couplings obtained from COHERENT (solid blue and dotted magenta) and CHARM + CDHS (dashed green) data. Top panels correspond to $\varepsilon_{e\tau}^{qV}$ and $\varepsilon_{e\mu}^{qV}$, while the bottom panel corresponds to $\varepsilon_{\mu\tau}^{qV}$.}
\label{Result:Vector:2} 
\end{figure}

For completeness, we also study the case of flavor-changing NSI parameters, with the corresponding results shown in Fig. \ref{Result:Vector:2}. The top-left panel shows the results for $\varepsilon_{e\tau}^{qV}$, with the COHERENT analysis shown in magenta (\textit{down} quark) and blue (\textit{up} quark). The CHARM result, which is blind to the quark type, is shown in green. We notice that in this case the COHERENT bounds are dominant, while CHARM shows poor sensitivity. By comparing our results, we notice that the COHERENT bounds are better for $down$ quark couplings than for $up$ quark ones, which contrasts with the scalar case presented above.  The top right panel in the same figure shows the sensitivity to $\varepsilon_{e\mu}^{qV}$. We can see that in this case, the sensitivity at 90\% C.L., i.e.,  $\Delta\chi^2 = 2.71$, is of the same order for both COHERENT and the combination CHARM+CDHS, where the latter does not distinguish between $up$ and $down$ quark couplings.  However, the sensitivity is slightly better for COHERENT.  Similarly, the bottom panel of Fig. \ref{Result:Vector:2} shows a comparable sensitivity between the two experiments for $\varepsilon_{\mu\tau}^{qV}$.  We now summarize the best obtained bounds in each case:

\begin{equation}
\begin{array}{rcll}
-0.146 \;<\; \varepsilon_{e\tau}^{uV} \;<\; 0.146
  &&& \text{(COHERENT)}
  \\[10pt]

-0.130 \;<\; \varepsilon_{e\tau}^{dV} \;<\; 0.130
  &&& \text{(COHERENT)}
  \\[10pt]

-0.080 \;<\; \varepsilon_{e\mu}^{uV} \;<\; 0.080
  &&& \text{(COHERENT )}
  \\[10pt]

-0.073 \;<\; \varepsilon_{e\mu}^{dV} \;<\; 0.073
  &&& \text{(COHERENT)}
  \\[10pt]

-0.096 \;<\; \varepsilon_{\mu\tau}^{uV} \;<\; 0.096
  &&& \text{(COHERENT)}
  \\[10pt]

-0.087 \;<\; \varepsilon_{\mu\tau}^{dV} \;<\; 0.087
  &&& \text{(COHERENT)}
\end{array}
\end{equation}

\begin{figure}[t]
\centering
\includegraphics[scale=0.7]{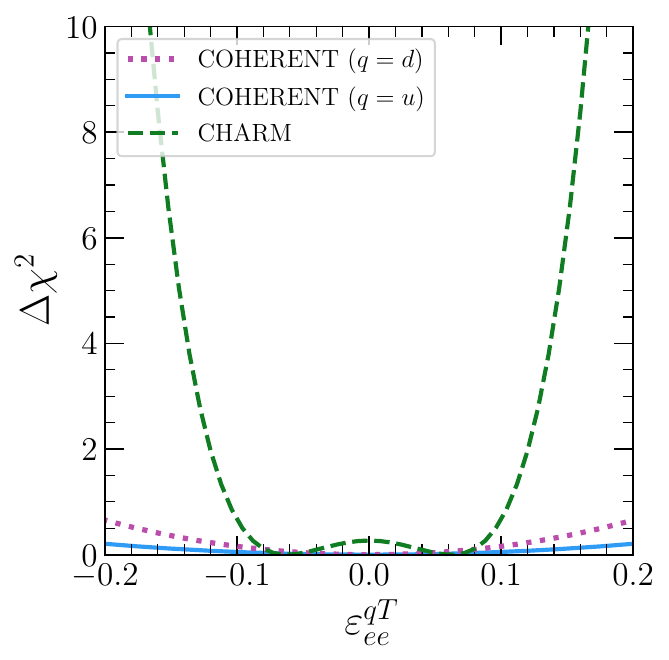}
\includegraphics[scale=0.7]{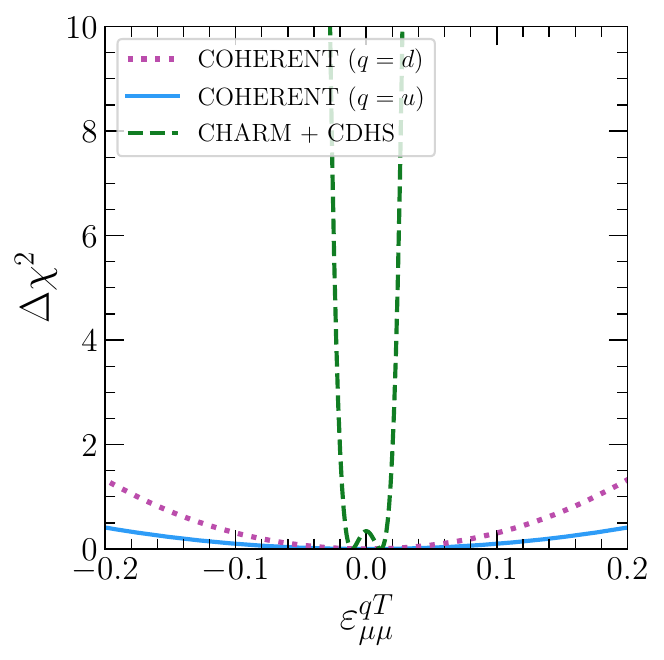}
\includegraphics[scale=0.7]{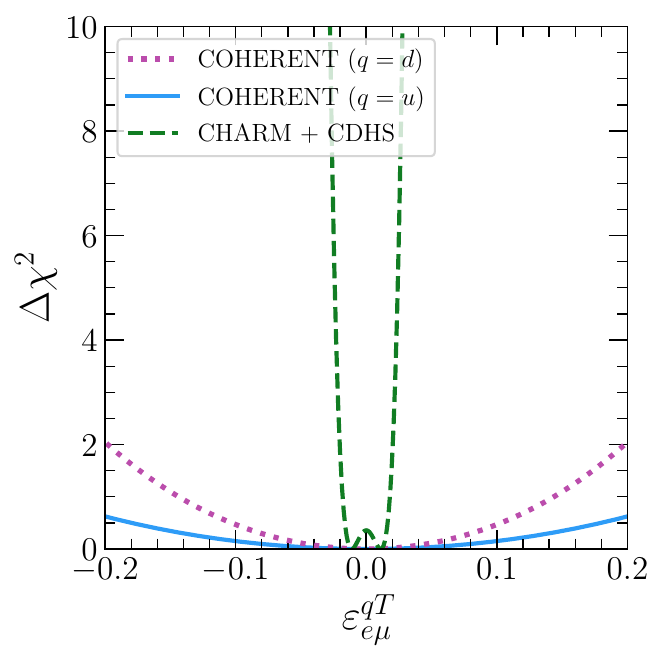}
\caption{$\Delta\chi^2$ profile for tensor interactions of up and down quark couplings obtained from COHERENT (solid blue and dotted magenta) and CHARM + CDHS (dashed green) data. Top panels correspond to $\varepsilon_{ee}^{qT}$ and $\varepsilon_{\mu\mu}^{qT}$, while the bottom panel corresponds to $\varepsilon_{e\mu}^{qT}$.}
\label{Result-Tensor} 
\end{figure}

To finalize the study of GNI parameters, we show in the panels of Fig.~\ref{Result-Tensor}, the results corresponding to  tensor interactions, assuming that there are no scalar nor vector contributions, and using the same color code as in the previous scenario.  The top left panel of the figure shows the results for $\varepsilon_{ee}^{qT}$. We notice that, even though they are of the same order, CHARM (dashed green) achieves the best sensitivity, while COHERENT shows poor sensitivity for both $up$ (solid blue) and $down$ (dotted magenta) quark couplings. The top right panel of the figure shows the corresponding results for $\varepsilon_{\mu\mu}^{qT}$. Again, we see that CHARM gives the best sensitivity. Moreover, in this case, the CHARM bounds are one order of magnitude better when compared to the case of COHERENT.  A similar conclusion is obtained for $\varepsilon_{\mu\mu}^{qT}$ from the bottom panel of the figure. The constraints from these  analyses are listed below

\begin{equation}
\begin{array}{rcll}
-0.122 \;<\; \varepsilon_{ee}^{qT} \;<\; 0.122
  &&& \text{(CHARM)}
  \\[10pt]

-0.021 \;<\; \varepsilon_{\mu\mu}^{qT} \;<\; 0.021
  &&& \text{(CHARM+CDHS)}
  \\[10pt]

-0.022 \;<\; \varepsilon_{e\mu}^{qT} \;<\; 0.022
  &&& \text{(CHARM+CDHS)}
\end{array}
\end{equation}

which shows that deep-inelastic scattering experiments are better at constraining tensor interactions.

\section{Conclusions}
After summarizing the two different parametrizations present in the literature for the study of GNI, we have presented in this paper a treatment of the parametrization that allows to perform a global analysis of the GNI parameters and a comprehensive guide to transform from one parametrization into another. 
We have studied the interplay between different neutrino scattering experiments to constrain GNI parameters.  In particular, we have considered the case of CEvNS using data from the COHERENT experiment and deep-inelastic scattering using a combined analysis of CHARM and CDHS data. Our results show a clear complementarity between these processes to study GNI. On the one hand, DIS experiments provide better constraints for tensor interactions, while CEvNS work better to constrain scalar parameters due to the enhancement of the associated cross sections. In the case of vector interactions, also known as NSI, the two processes provide similar results.  
\label{sec:conclusions}

\acknowledgements 
\noindent We thank Martín González-Alonso for fruitful discussions on tensor interactions. This work has been supported by SNII
(Sistema Nacional de Investigadoras e Investigadores, Mexico). L. J. F. has been supported by the project SECIHTI CBF-2025-I-1589 and DGAPA UNAM grant
PAPIIT IN111625. 
\bibliographystyle{utphys}
\bibliography{coherent}

\end{document}